\documentclass[aps,prl,reprint,superscriptaddress,longbibliography]{revtex4-2}
\usepackage[T1]{fontenc}
\usepackage{amsmath,amssymb,bm,graphicx,xcolor}
\usepackage[colorlinks=true,citecolor=blue,linkcolor=blue,urlcolor=blue]{hyperref}
\begin{document}
\title{Broadband molecular dynamics simulation of confined Brownian motion}
\author{Quentin Thomas}
\affiliation{Laboratoire Interfaces \& Fluides Complexes, Universit\'{e} de Mons, 20 Place du Parc, B-7000 Mons, Belgium}
\author{Clara Mefo Sop}
\affiliation{Laboratoire Interfaces \& Fluides Complexes, Universit\'{e} de Mons, 20 Place du Parc, B-7000 Mons, Belgium}
\author{Maxime Lavaud}
\affiliation{Univ. Bordeaux, CNRS, LOMA, UMR 5798, 33405 Talence, France}
\affiliation{Faculty of Physics, University of Vienna, Boltzmanngasse 5, 1090 Vienna, Austria}
\author{Yacine Amarouchene}
\email{yacine.amarouchene@u-bordeaux.fr}
\affiliation{Univ. Bordeaux, CNRS, LOMA, UMR 5798, 33405 Talence, France}
\author{Thomas Salez}
\email{thomas.salez@cnrs.fr}
\affiliation{Univ. Bordeaux, CNRS, LOMA, UMR 5798, 33405 Talence, France}
\affiliation{Mechanics Department, Ecole Polytechnique, Institut Polytechnique de Paris, 91128 Palaiseau, France }
\author{Pascal Damman}
\email{pascal.damman@umons.ac.be}
\affiliation{Laboratoire Interfaces \& Fluides Complexes, Universit\'{e} de Mons, 20 Place du Parc, B-7000 Mons, Belgium}
\date{\today}
\begin{abstract}
Molecular dynamics with an explicit solvent resolves the hydrodynamic response of a Brownian colloid in bulk and near a rigid wall over acoustic, inertial, and diffusive time scales. Bulk spectra establish a continuum reference and reveal departures associated with discrete-solvent dynamics. Under confinement, thermal spectra provide a measurement of the space-dependent added mass, while velocity autocorrelations determine space-dependent, anisotropic  hydrodynamic memory and diffusion. All together, these measurements connect the enhanced added mass and reduced mobility near a wall, and probe the discrete-continuum crossover.
\end{abstract}
\maketitle

Brownian motion is one of the paradigmatic manifestations of molecular thermal fluctuations and has played a central role in the development of statistical physics since the pioneering works of Einstein, Smoluchowski, and Langevin~\cite{duplantier2005mouvement,bian2016}. The classical Langevin description assumes a memoryless process related to the instantaneous viscous response of the surrounding fluid allowing the dissipation of the particle momentum. This approximation leads to the Einstein relation between the diffusion coefficient and the hydrodynamic friction coefficient. 
However, Brownian motion is not purely diffusive. Momentum transferred from a particle to the surrounding fluid is transported through the latter and subsequently feeds back onto the particle motion. This hydrodynamic memory gives rise to algebraic long-time tails in the velocity autocorrelation function (VACF), as first revealed by Alder and Wainwright~\cite{Alder1970} and later described by fluctuating and unsteady hydrodynamics~\cite{Zwanzig1970,Zwanzig1975,Hinch1975,Clercx1992,LiRaizen2013}.  Modern experiments have directly observed this memory and resolved the crossover between inertial and diffusive regimes~\cite{li2010measurement,Franosch2011,kheifets2014}.
Moreover, at very short times, acoustic propagation controls the initial transfer of momentum, preceding the effectively incompressible response~\cite{Zwanzig1975,LiRaizen2013}.

Boundaries enrich this physics by reorganizing the flow due to the boundary conditions. The mobility of a Brownian particle near a rigid wall is known to be anisotropic and inhomogeneous, with diffusion coefficients parallel and normal to the interface differing from their bulk values~\cite{Brenner1961,Faucheux1994,Lin2000,Jeney2008,Mo2015,Lavaud2021,Millan2023}, and associated non-Gaussianities in displacements~\cite{Alexandre2023,Tournemire2026}. However, diffusion probes only the zero-frequency limit of the response. The high-frequency limit, where the inertia of the accelerated fluid induces an added mass and triggers memory effects, is harder to access~\cite{zhang2023unsteady,ferreira2026quantitative}, because it requires resolving the particle motion at short times. 

In this Letter, we show that molecular dynamics (MD) simulations with an explicit solvent allow to resolve the full temporal window of Brownian motion. We find that confinement not only changes the mobility, but the full hydrodynamic memory from the acoustic to the diffusive regimes. In practice, we analyze thermal fluctuations spectrally to infer the hydrodynamic force response through the fluctuation--dissipation relation. Bulk measurements first calibrate the continuum limit and the crossover to solvent-collision dynamics. The main study then resolves the space-dependent added mass, directional mobility, and hydrodynamic memory near a rigid wall, using velocity autocorrelations and trapped-particle spectra as complementary probes.

\begin{figure}[t]
\centering
\includegraphics[width=0.78\columnwidth]{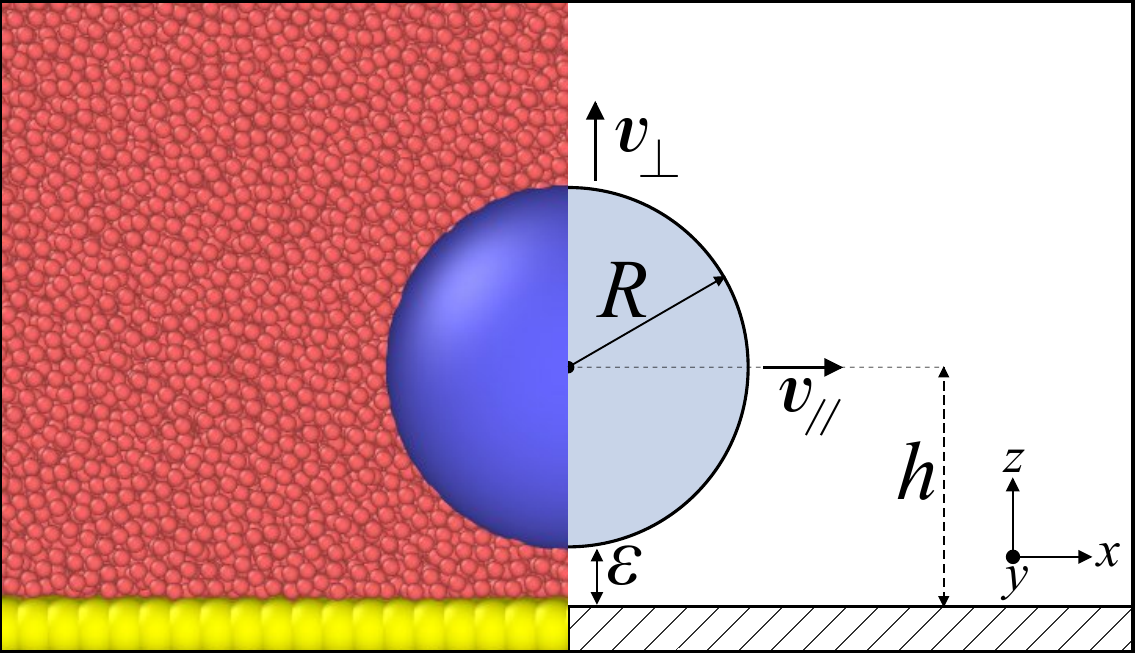}
\caption{Left: MD snapshot of a colloid in an explicit solvent near a rigid wall. Right: schematic defining the radius $R$, the surface gap $\epsilon$, the center-to-wall distance $h=R+\epsilon$, and the velocity components $v_\parallel$ and $v_\perp$.}
\label{fig:geometry}
\end{figure}

We simulate a neutrally-buoyant sphere in a dense Weeks--Chandler--Andersen solvent using LAMMPS~\cite{thompson2022lammps}. In reduced Lennard--Jones units, solvent particles have a mass $m=1$ and a diameter $\sigma=0.6$. Besides, the thermal energy is $k_{\mathrm B}T=750$ and the fluid density is $\rho_{\mathrm{f}}=3.3$. The smooth central colloid--solvent interaction is purely repulsive, with range $R+\sigma/2$. The associated force acts normally to the colloidal surface. The viscosity $\eta$ and sound speed $c_{\mathrm{s}}$ are measured independently~\cite{SM}, defining $\nu=\eta/\rho_{\mathrm{f}}$, $\tau_{\mathrm{f}}=R^2/\nu$, and $\tau_{\mathrm{c}}=R/c_{\mathrm{s}}$. Bulk simulations use periodic boundary conditions. Confined simulations further introduce a planar wall, as shown in Fig.~\ref{fig:geometry}. Simulation and sampling details are provided in the Supplemental Material (SM)~\cite{SM}.

\begin{figure*}[t]
\centering
\includegraphics[width=0.35\textwidth]{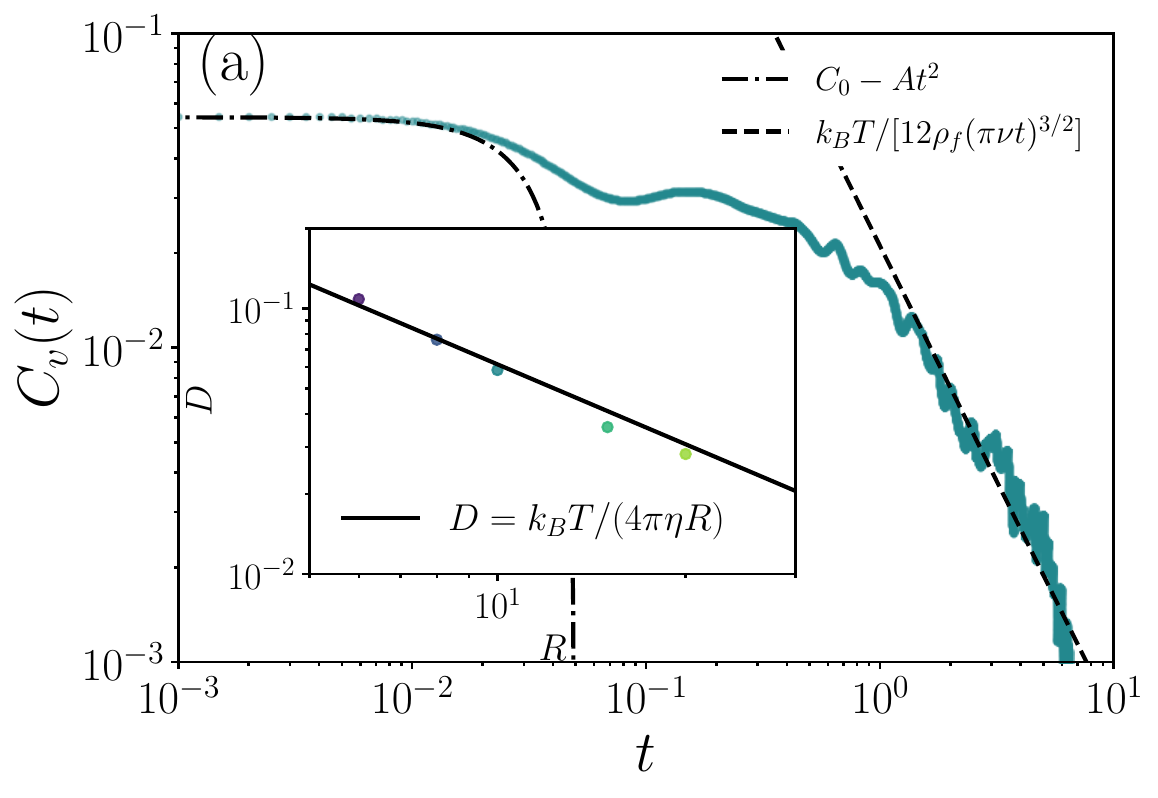}
\includegraphics[width=0.62\textwidth]{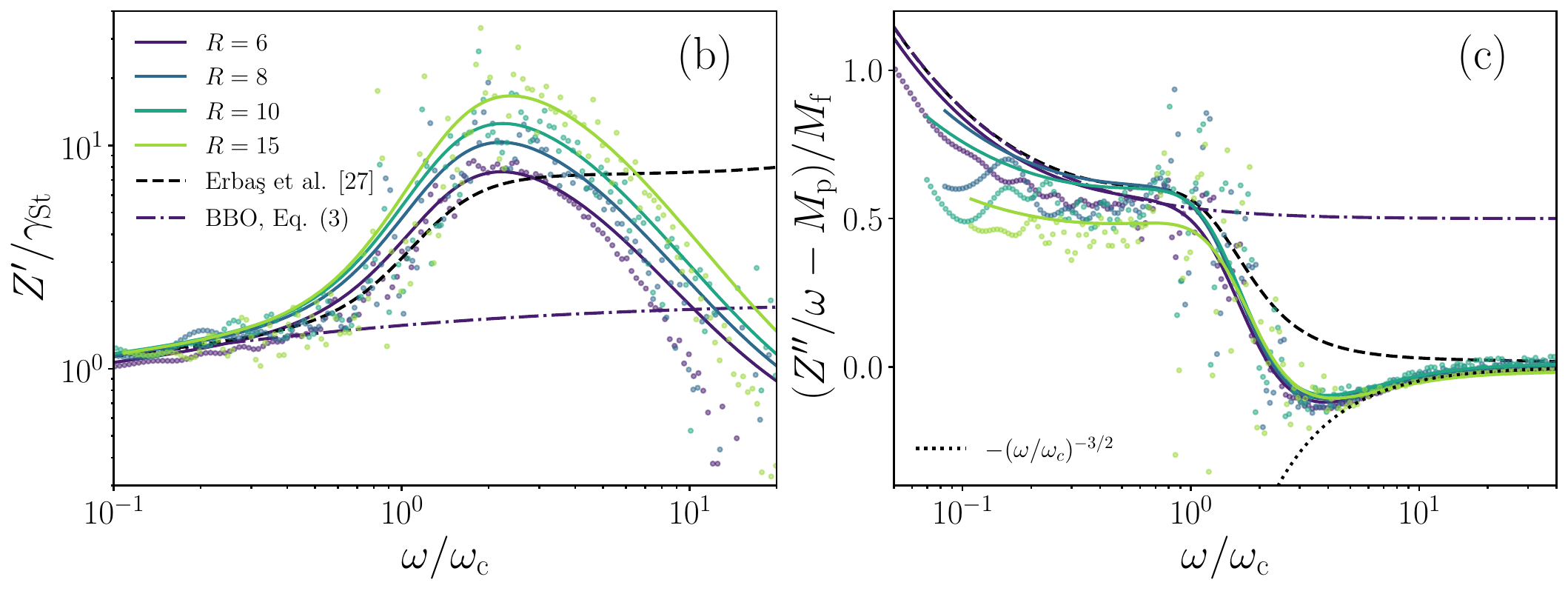}
\caption{Bulk calibration. (a) Velocity autocorrelation function $C_{\textrm{v}}$ as a function of time $t$ for a radius $R=10$. The theoretical long-time tail (see Eq.~\eqref{eq:bulk_tail}) and the short-time parabolic dynamics are also shown. Inset: evolution of the diffusion coefficient $D$ with the particle radius $R$. The solid line corresponds to  $D=k_{\textrm{B}}T/(4\pi\eta R)$. Dissipative (b) and inertial (c) responses computed from the VACFs versus angular frequency normalized by $\omega_{\textrm{c}}=1/\tau_{\textrm{c}}$, for $R=6,8,10,15$. Symbols are MD data. Dotted curves correspond to  Eq.~\eqref{eq:slip_supp} ($R=6$), while long-dashed curves correspond to the compressible-slip model~\cite{Erbas2010} ($R=6$), and solid curves to the compressible-slip model with the collisional transition (see Eq.~\eqref{eq:filtered}). Finally, short-dashed curves correspond to the high-frequency inertial response $-\omega^{-3/2}$. 
}
\label{fig:bulk}
\end{figure*}

\textit{Bulk dynamics.} From the MD trajectories, we compute the velocity autocorrelation functions $C_{v_i}(t)=\langle v_i(t)v_i(0)\rangle$, where $v_i$ is the $i$th Cartesian component ($i=x,y,z$) of the particle velocity $\mathbf v$. Bulk isotropy gives the same $C_{v_i}$, that we note $C_v$, for all three components. As shown in Fig.~\ref{fig:bulk}(a), it exhibits a complex behaviour with different regimes, including the long-time tail associated with viscous memory that has the classical bulk signature:
\begin{equation}
C_v(t)\simeq\frac{k_{\mathrm B}T}{12\rho_{\mathrm{f}}(\pi\nu t)^{3/2}}\ .
\label{eq:bulk_tail}
\end{equation} 
As shown in the inset of Fig.~\ref{fig:bulk}(a), the bulk diffusion coefficient estimated by the Green-Kubo method is consistent with $D_0=k_{\mathrm B}T/\gamma$, with $\gamma=4\pi\eta R$. This is the perfect-slip limit~\cite{Erbas2010,Theers2016}, in agreement with the smooth central colloid--solvent interaction that cannot transfer tangential forces. 

Furthermore, the fluctuation-dissipation relation gives the complex velocity impedance as:
\begin{equation}
\begin{aligned}
Z(\omega)&=\frac{k_{\mathrm B}T}{\widetilde C_v(\omega)}\ ,\\
\widetilde C_v(\omega)&=\int_0^\infty\mathrm dt\, C_{v}(t)\exp(\mathrm{i}\omega t)\ .
\end{aligned}
\label{eq:impedance}
\end{equation}
We use the convention $Z=Z'-\mathrm{i}Z''=-\mathrm{i}\omega M_{\mathrm{p}}+G(\omega)$, so that a positive inertial contribution appears in $Z''$. The fluid response $G$ relates an imposed velocity component to the corresponding fluid force, $F_{\mathrm{hyd},i}(\omega)=-G(\omega)v_i(\omega)$. The real part $Z'$ measures dissipative contributions, while $Z''/\omega-M_{\mathrm{p}}$ measures the additional fluid contribution to the inertial response, which is expected to scale like the displaced fluid mass $M_{\mathrm{f}}=4\pi\rho_{\mathrm{f}}R^3/3$. For the Boussinesq-Basset-Oseen (BBO) generalized friction force, the hydrodynamic memory contributes to both the real and imaginary parts of $Z$. Figure~\ref{fig:bulk}(b,c) presents the impedance obtained for several radii from the MD data, using  Eq.~\eqref{eq:impedance} and the acoustic angular frequency $\omega_{\mathrm{c}}=1/\tau_{\mathrm{c}}$ to normalize the angular frequency. 
The perfect-slip BBO force~\cite{Erbas2010,Mo2017} is given by: 
\begin{equation}
G_{\mathrm{slip}}/\gamma_\mathrm{st}=\frac 23
+\frac 43\left(\frac{s}{3+s}\right)-\frac{\mathrm{i}\omega M_{\mathrm{f}}}{2\gamma_\mathrm{st}},\ s=\sqrt{-\mathrm{i}\omega\tau_{\mathrm{f}}}\ ,
\label{eq:slip_supp}
\end{equation}
where $\gamma_\mathrm{st}=6\pi\eta R$ is the no-slip friction coefficient. 
It agrees with the data in the low-frequency range, where hydrodynamic memory dominates. Besides, the added mass $M_{\mathrm{a}}$ is defined as the fluid contribution to the effective particle mass, $M^*=M_{\mathrm{p}}+M_{\mathrm{a}}$. In the incompressible inertial regime ($\omega\tau_{\mathrm{f}}\gg1$ and $\omega\tau_{\mathrm{c}}\ll1$), potential-flow theory gives $M_{\mathrm{a}}=M_{\mathrm{f}}/2$~\cite{Lamb1932}, hence ${(Z''/\omega-M_{\mathrm{p}})}/{M_{\mathrm{f}}} = 1/2$. The data shown in Fig.~\ref{fig:bulk}(c) does exhibit this regime at intermediate frequencies. For $\omega\to 0$, the dissipative component reaches the expected limit of the BBO expression, $Z'/\gamma_\mathrm{st} \simeq 2/3$. We stress that, for $\omega\tau_{\mathrm{c}}\ll1$, acoustic relaxation is fast compared with the colloidal imposed motion and the fluid is effectively incompressible. 

When $\omega\tau_{\mathrm{c}}$ becomes comparable to unity, deviations from these predictions are observed: $Z'/\gamma_\mathrm{st}$ increases well above the BBO limit and a drop of $Z''/\omega-M$ is observed. In this frequency range, compressibility becomes important. We therefore compare the data with the compressible perfect-slip response proposed by Erba\c{s}, Podgornik, and Netz~\cite{Erbas2010}. It provides the continuum reference for the acoustic crossover through the relation $Z_{\mathrm{model}}(\omega)=-\mathrm{i}\omega M_{\mathrm{p}}+ G_{\mathrm{comp}}(\omega)$,  where the expression of $G_{\mathrm{comp}}$ is given in~\cite{SM}. The prediction of this model captures well the data in the low-frequency to moderate-frequency range. In particular, the bare mass is recovered as $(Z''/\omega - M_{\mathrm{p}})$ approaches 0, consistent with the energy equipartition, $C_{\mathrm{v}}(0)=k_{\mathrm B}T/M_{\mathrm{p}}$~\cite{Zwanzig1975}. 

At frequencies above the acoustic crossover, $\omega\tau_{\mathrm c} > 1$, the measured impedance deviates markedly from the compressible perfect-slip prediction, displaying a decrease of $Z'/\gamma_\mathrm{st}$ and a negative fluid contribution $Z''/\omega-M_{\mathrm p}$ to the effective mass. These deviations suggest that a continuum description with frequency-independent viscosities becomes insufficient and that molecular-scale solvent relaxation must be considered.
The earliest-time VACF in Fig.~\ref{fig:bulk}(a) is well described by $C_v(t)\simeq C_v(0)-At^2$~\cite{Boon1980}. The observed vanishing initial slope is consistent with the regular forces generated by the WCA interactions. This behavior differs from a purely exponential Enskog VACF corresponding to an instantaneous collisional friction. Instead, the evolution of the VACF is compatible with a short-time force memory arising from finite-duration molecular encounters, although it does not uniquely identify its functional form~\cite{Rickayzen2005}.
Therefore, to describe the impedance over the full frequency range, we consider the following ansatz: 
\begin{equation}
Z_{\mathrm{model}}(\omega)
=-\mathrm{i}\omega M_{\mathrm p}
+\frac{G_{\mathrm{comp}}(\omega)}
{1-\mathrm{i}\omega\tau_{\mathrm{mol}}}\ .
\label{eq:filtered}
\end{equation} 
Such a low-pass-filter structure preserves the steady friction and introduces an additional relaxation time scale $\tau_{\mathrm{mol}}$, representing an effective delay in the hydrodynamic force response. It phenomenologically accounts for short-time effects beyond the compressible Newtonian continuum description, potentially involving finite collision durations and solvent rearrangements near the particle. As shown in Fig.~\ref{fig:bulk}(b,c), such an ansatz captures the measured dissipative and inertial responses over the considered frequency range. 
In particular, in the large-frequency asymptotic limit where $G_{\mathrm{comp}}\propto\sqrt{-\mathrm{i}\omega}$, the ansatz predicts $Z''/\omega-M_{\mathrm p}\propto-\omega^{-3/2}$, which is consistent with the data. 
Interestingly, the fitted memory time $\tau_{\mathrm{mol}}$ appears to scale with the colloidal radius~\cite{SM}. This additional relaxation may reflect a transient elastic response of the solvent layer surrounding the colloid, arising from motion-induced local deformations and subsequent molecular rearrangements near its surface.  

\textit{Confined dynamics.}
Let us now consider a planar rigid wall and investigate the space-dependent and anisotropic dynamics of the particle nearby. We introduce the $j=\parallel,\perp$ directions and the gap distance $\epsilon$. We use the following setup: for parallel motion, the particle height is constrained while the measured in-plane coordinate is free; for normal motion, the particle is equilibrated at the selected gap and released; velocity autocorrelations are sampled before substantial gap evolution. 
The dynamical response is then measured using $C_j(t)$ and Eq.~\eqref{eq:impedance}, like in the bulk case. 

\begin{figure}[t]
\centering
\includegraphics[width=0.8\columnwidth]{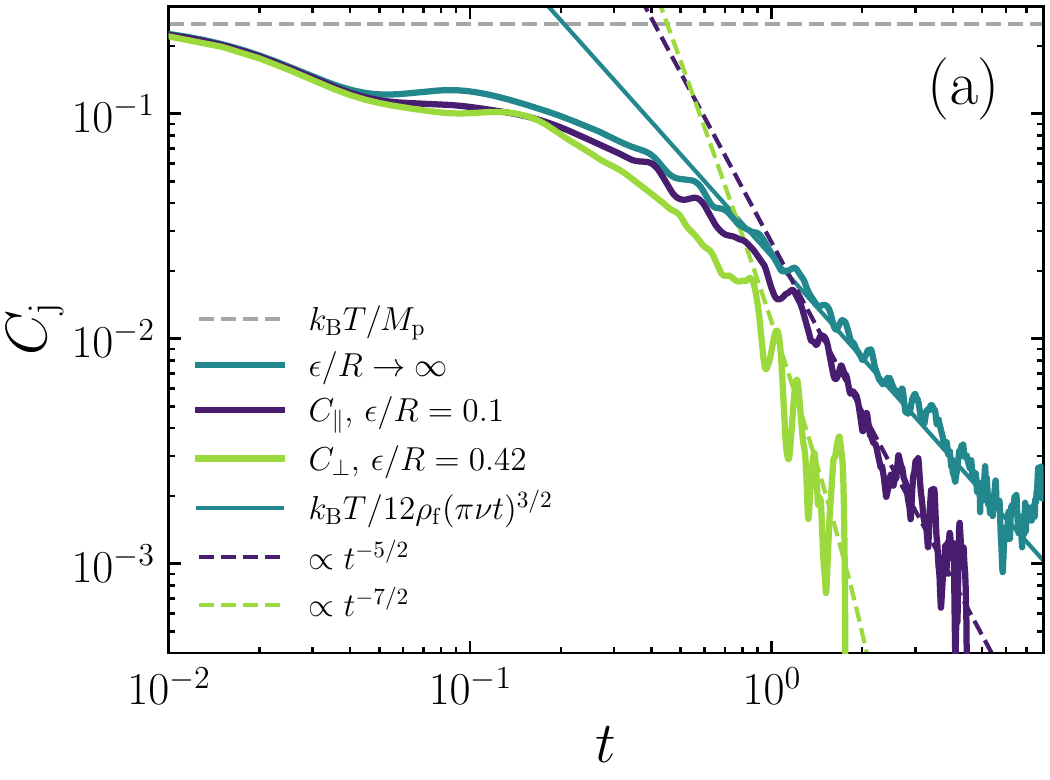}
\includegraphics[width=0.8\columnwidth]{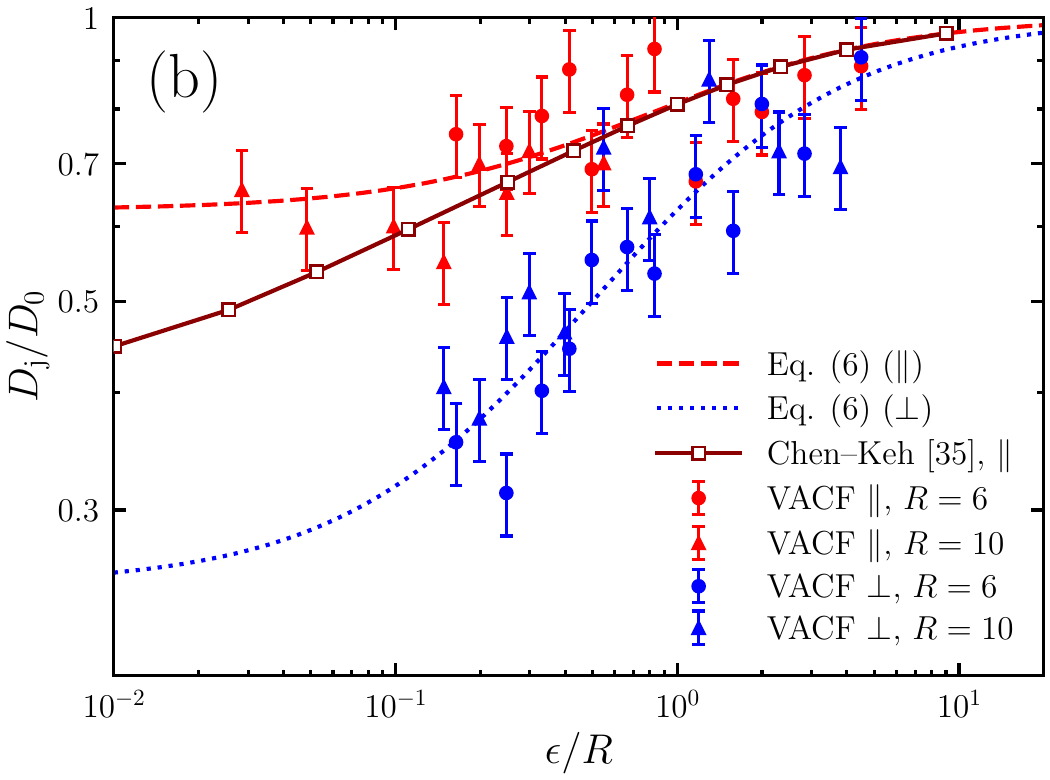}
\caption{(a) VACFs $C_i$ (with $i\in \{\parallel, \perp\}$) as functions of time $t$, for bulk and confined situations. The bulk long-time tail (see Eq.~\eqref{eq:bulk_tail}) and power-law fits for parallel and perpendicular long-time tails in confinement are shown with dashed lines. (b) Normalized local diffusion coefficients as functions of confinement parameter $\epsilon/R$, for $R=6$ and $10$, computed (symbols) either from Eq.~\eqref{eq:diffusion} or from low-frequency PSDs. The solid curve with squares shows the numerical data from~\cite{chen2003}. The dashed and dotted lines indicate the asymptotic relations for parallel and perpendicular motions~\cite{procopio2024}, respectively.}
\label{fig:wall}
\end{figure}

Figure~\ref{fig:wall}(a) shows that the VACFs of confined particles decay faster than in the bulk at long time, and split into distinct parallel and normal responses. The respective long-time tails scale as $t^{-5/2}$ and $t^{-7/2}$. These exponents originate from the long-wavelength diffusion and reflection of vorticity at the wall and are therefore expected to remain unchanged by slippage at the particle surface. The latter nevertheless modifies the coupling between the particle and the surrounding fluid and hence the amplitudes of these algebraic tails. As such, the prefactors derived by Felderhof for a no-slip particle near a no-slip wall~\cite{Felderhof2005} cannot be directly used here and, to our knowledge, their counterparts for a perfect-slip sphere near a no-slip wall are not presently available. 
The steady mobility can be obtained from the Green-Kubo relation:
\begin{equation}
D_j(\epsilon)=\int_0^\infty\mathrm dt\, C_j(t,\epsilon)
=\frac{k_{\mathrm B}T}{Z_j(0,\epsilon)}\ .
\label{eq:diffusion}
\end{equation}
As previously observed for no-slip spheres, the diffusion coefficients of a perfect-slip particle decrease in confinement (Fig.~\ref{fig:wall}). 
In the far-field limit ($\epsilon\gg R$), the diffusion coefficients can be obtained from the hydrodynamic reflection method developed for a sphere with arbitrary Navier slip near a no-slip wall \cite{procopio2024}. Taking the perfect-slip limit yields:
\begin{equation}
\frac{D_\parallel}{D_0}
\simeq 1-\frac{3R}{8\epsilon}\ ,
\qquad
\frac{D_\perp}{D_0}
\simeq 1-\frac{3R}{4\epsilon}\ .
\end{equation}
As shown in Fig.~\ref{fig:wall}(b), the diffusion coefficients computed from the VACF are in good agreement with these asymptotic theoretical laws and with the numerical derivation of the parallel diffusion coefficients by Chen and Keh~\cite{chen2003}.

\begin{figure*}[t]
\centering
\includegraphics[width=\textwidth]{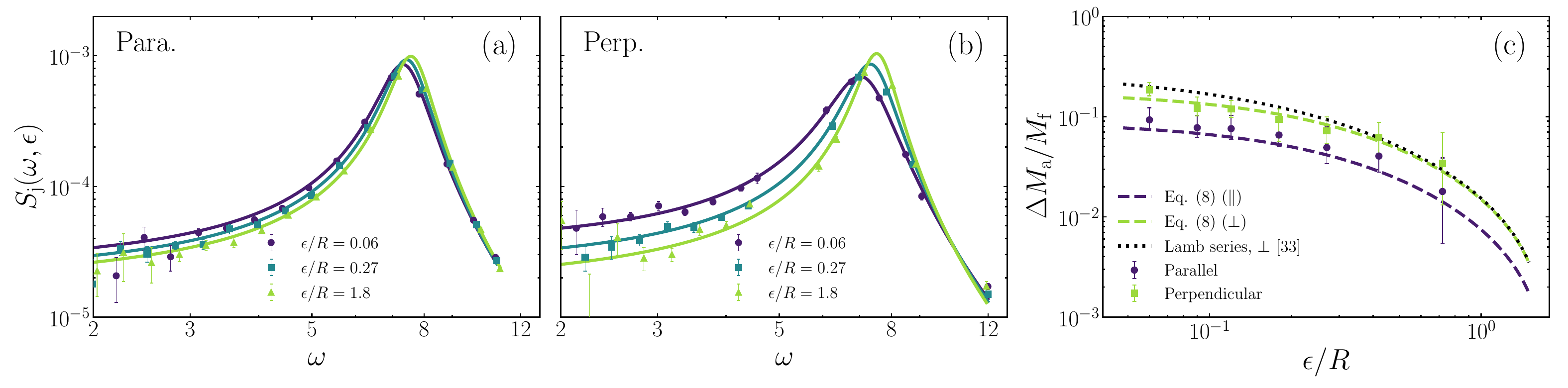}
\caption{Positional power spectral densities $S_j$ of trapped colloids as functions of angular frequency $\omega$, for various confinement parameters $\epsilon/R$, as indicated: (a) parallel direction ; (b) perpendicular direction. Symbols are MD simulation results and solid lines are fits to Eq.~\eqref{eq:psd}, with the fluid response specified in the SM~\cite{SM}. (c) Normalized confinement-induced increase in fluid added-mass contributions $\Delta M_{\mathrm{a},j}/M_{\mathrm{f}}$ (see definition in main text), as functions of confinement parameter $\epsilon/R$, for both the parallel and perpendicular directions. Symbols are MD simulation results. The solid and dashed lines correspond to the normal-motion Lamb series and to the leading-order potential-flow theoretical predictions, respectively.}
\label{fig:mass}
\end{figure*}

Regarding the added mass, the impedances $Z$ computed for the confined particles are unfortunately too noisy near the wall for a reliable separation of added-mass and memory effects. To have a complete picture of the dynamics, we therefore access this part of the response through the thermal positional spectra of a colloid in a harmonic trap of known stiffness $K$.
For a trapped particle, the two-sided positional power-spectral density (PSD) reads:
\begin{equation}
S_{j}(\omega,\epsilon)=
\frac{2k_{\mathrm B}T\operatorname{Re}G_j(\omega,\epsilon)}
{|K-M_{\mathrm{p}}\omega^2-\mathrm{i}\omega G_j(\omega,\epsilon)|^2}\ .
\label{eq:psd}
\end{equation}
A direct examination of the PSDs in Fig.~\ref{fig:mass}(a,b) reveals interesting features about confinement. When the gap diminishes, we observe a decrease of the spectral resonance frequency and an increase of the low-frequency plateau that highlight the increase of the added mass and the extra dissipation induced by the confinement. These effects are more pronounced for the perpendicular motion than for the parallel one. We fit the PSDs using the perfect-slip incompressible BBO response [Eq.~\eqref{eq:slip_supp}] for $G_j$, with anisotropic space-dependent stationary drag coefficients $\gamma_j(\epsilon)$ and added-mass contributions $-\mathrm{i}\omega M_{\mathrm{a},j}(\epsilon)$. The agreement is good, as shown in Fig.~\ref{fig:mass}(a,b). Below, we discuss the extracted fit parameters $\gamma_j(\epsilon)$ and $M_{\mathrm{a},j}(\epsilon)$.
As shown in Fig.~\ref{fig:mass}(c), the shift in added mass, $\Delta M_{\mathrm{a},j}=M_{\mathrm{a},j}(\epsilon/R)-M_{\mathrm{a},j}(\infty)$ induced by the confinement for parallel and perpendicular motion follow the prediction of the potential-flow theory~\cite{Simha2018}, 
\begin{equation}
\frac{\Delta M_{\mathrm{a},j}}{M_{\mathrm{f}}}\simeq a_j
\left(\frac{R}{R+\epsilon}\right)^3,
\label{eq:reflection}
\end{equation}
where $a_\parallel=3/32$ and $a_\perp=3/16$. For perpendicular motion, the data also agree with the higher-order corrections following the Lamb--Milne-Thomson series~\cite{Lamb1932,Yang2010}. 

Finally, the diffusion coefficients of Brownian particles near a wall can be also obtained from the fitting of the PSD through the Einstein relation, $D_j(\epsilon)=k_\mathrm BT/\gamma_j(\epsilon)$. These additional estimates are consistent with the diffusion coefficients calculated using the Green-Kubo integration of the VACF.

In conclusion, explicit-solvent MD simulations provide broadband access to the dynamics of bulk and confined colloids. Diffusion in the bulk shows that the simulated colloidal particles interact with the fluid through a perfect-slip boundary condition. In the high frequency range, a crossover from the continuum regime, including both inertial and dissipative effects, to a collisional regime is observed and rationalized. Near a rigid wall, the same spectral framework allows to unify the enhanced, space-dependent, anisotropic fluid added mass to the hindered, space-dependent, anisotropic  mobility, including the modified hydrodynamic memory. As such, our combined analysis verifies key hydrodynamic predictions in confinement, and explores the molecular breaking of continuum hydrodynamics.

\begin{acknowledgments}
The authors thank Vincent Bertin, Quentin Ferreira, Thomas Franosch, Harshit Joshi, Juliette Lacherez and Marco Rizzo for interesting discussions. Quentin Thomas is a FRIA grant holder from the Belgian National Fund for Scientific Research (FRS-FNRS). This work was supported by the F.R.S.-FNRS under the research Grant (PDR ``Active Matter in Harmonic Trap'') No. T.0251.20, by the European Union through the European Research Council under EMetBrown (ERC-CoG-101039103) grant, and by the Interdisciplinary and Exploratory Research program under MISTIC grant at the University of Bordeaux. The Belgian National Fund for Scientific Research (FRS-FNRS) within the Consortium des Equipements de Calcul Intensif – CECI, under Grant 2.5020.11t is also acknowledged. Computer time for this study was provided by the computing facilities of the M\'esocentre de Calcul Intensif Aquitain.
The authors thank the Soft Matter Collaborative Research Unit, Frontier Research Center for Advanced Material and Life Science, Faculty of Advanced
Life Science at Hokkaido University, Sapporo, Japan, and the CNRS International Research Network between France and India on Hydrodynamics at small scales:
From soft matter to bioengineering.
\end{acknowledgments}

\end{document}